\documentclass{icrc2009}

\usepackage{graphicx}   
\usepackage{caption}    
\usepackage[font=footnotesize]{subfig} 
\usepackage{fixltx2e}
\usepackage{url}
\usepackage{amssymb}
\usepackage{lineno}

\newcommand{\shorttitle}[1]%
{\markboth{Proceedings of the 31\MakeLowercase{$^{st}$} ICRC, {\L}\'{o}d\'{z} 2009}{#1} }
\newcommand{\etal}{\MakeLowercase{\textit{et al. }}} 

\begin{document}
\title{Studying individual UHECR sources with high statistics}

\author{\IEEEauthorblockN{Medina-Tanco G.\IEEEauthorrefmark{1},
	for the JEM-EUSO Collaboration}
                            \\
\IEEEauthorblockA{\IEEEauthorrefmark{1}Instituto de Ciencias Nucleares (ICN), Universidad Nacional Aut\'onoma de M\'exico, M\'exico D.F.}
}

\shorttitle{Medina-Tanco G. \etal Studying individual UHECR sources with high statistics}
\maketitle

\begin{abstract}
One of the main impacts of the JEM-EUSO mission will come from its 
unprecedented exposure. This feature creates, for the first time 
in the field, the possibility of studying individual UHECR sources. 
However, the intrinsic characteristics of the sources and the injection 
mechanism, as well as the presence of intervening magnetic fields, 
present challenges to the identification of isolated sources and to 
the extraction of their relevant spectral information from the data. 
We discuss here these aspects in a quantitative way and give an overview 
of what can be achieved in this regard under different astrophysical 
scenarios.  

\end{abstract}


 
\section{Introduction} \label{sec:intro}

The JEM-EUSO mission \cite{Toshi2009} will attain along its lifetime an unprecedented 
exposure of $\gtrsim 10^{6}$ km$^{2}$ sr yr. Depending on the assumptions made on 
the cosmic ray energy spectrum such exposure could translate, in practice, in 
$\sim 5 \times 10^{3}$ events above $10^{19.7}$ eV \cite{Inoue2009}. Since, conservatively, the ultra-high energy
cosmic ray (UHECR) flux must be dominated by relatively near-by sources, of which few 
are expected, it is very likely that high multiplicity clusters of events will be detected by 
JEM-EUSO in association with those sources. The observation of such clusters opens 
the first real possibility of identifying individual astrophysical objects responsible for 
UHECR production.  

The possibility of individual source detection, however, rests on the as yet unknown 
structure and intensity of the intervening magnetic fields. UHECR are almost certainly 
extragalactic and, therefore, have to traverse different environments with peculiar magnetic
structures in their way to the Earth: (i) the immediate magnetize envelope of the source, 
(ii) the intergalactic medium, (iii) the Galactic Halo and (iv) the Galactic disk. Our knowledge
of these regions is increasingly poorer as we move away from Earth and the Galaxy and,
even in the case of the Galactic magnetic field (GMF), the best understood region is our
local vicinity up to scales of few kpc, where most pulsars with known distances are located
and where the intervening interstellar medium can be acceptably probed. 

Few measurements exists of the intergalactic magnetic field (IGMF) and most of them
are limited to upper limits coming from Faraday rotation measurements, which establish that
$B \times L_{c}^{1/2}  \leqq 10^{-9}$ G Mpc$^{1/2}$ \cite{Kronberg94}. Since the actual value of the correlation
length $L_{c}$ is unknown, the later upper limit is not very restrictive and different scales can 
accommodate, in principle, very different magnetic field intensities. In fact, if one assumes
that the IGMF derives from the injection of magnetic field by primeval dwarf Galaxies
around $z=10$, then the magnetic filling factor at $z=0$ can vary between $\sim 10$ and 
$\sim 80$ \% depending on the chosen boundary conditions \cite{Kronberg2002}. Thus, 
at present, it is not at all clear what is the structure of the IGMF, which can range from
a highly spatially heterogeneous distribution, with magnetic intensities of $\sim 0.1 -1.0$
$\mu$G inside cosmic filaments and walls and negligible values inside large voids of 
large $L_{c}$, to large filling factor fields which take more moderate values
both in voids and high density regions ($\sim 10^{-10} - 10^{-8}$ G). These two regimes 
are very different from the point of view of UHECR propagation. The large filling factor regime 
allows for a quasi-ballistic propagation with small lost of directional information, while 
the low filling factor scenario probably leads to diffusion and drifts along walls and 
filaments, destroying all correlation between the sources location on the sky and the
arrival directions of particles to the detector. 

The Halo magnetic field is largely unknown despite the fact that it can be a very 
important factor in determining our ability to correlate UHECR events with individual 
sources. Depending on the mechanical energy output of our Galaxy, the magnetized 
Halo can vary widely in physical extent and degree of structuring. Its effects can 
range consequently from focusing and amplification due to the presence of caustics, 
to mild spreading of the pointing directions at the external border of the Halo.

Furthermore, recent results from Auger \cite{AugerScience2007} find a maximum 
correlation with objects of the Veron-Cetty-Veron catalog for a search radius of $3.2$ 
deg and source distances of $\sim 70$ Mpc at $E \gtrsim 60 EeV$. However, the 
trajectory bending at these energies due to the GMF already amounts to a few degrees. 
On the other hand, if we approximate the propagation of the particle as a random walk,
which is strictly valid only in the large filling factor scenario, the deflection can be written 
as:

\begin{equation}
\delta \theta = 0.42^{o} Z \left( \frac{E}{60 EeV} \right) \left( \frac{D}{L_{c}} \right)
                     \left(  \frac{B}{10^{-9} G} \right)
\end{equation}

\begin{figure}[!t]
\includegraphics [width=0.48\textwidth]{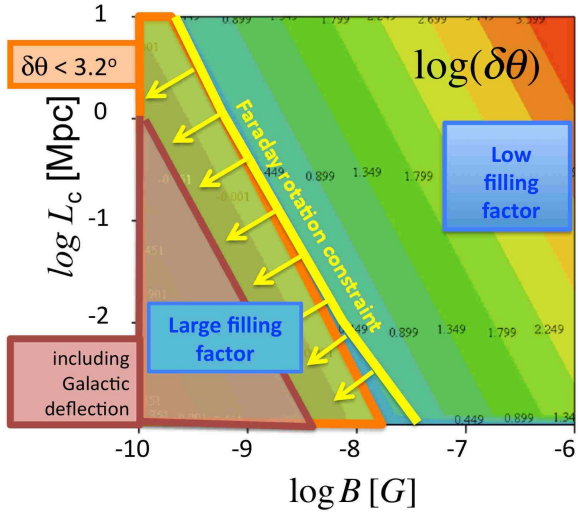}
\caption{UHECR deflection in the IGMF (log scale contour levels) as a function of the 
correlation length and the average intensity of the magnetic field. An average deflection $\delta \theta < 3.2^{o}$ basically rules out low filling factor IGMF models.}
\label{IGMFUpperLimit}
\end{figure}

Therefore, if the optimal correlation radius can be interpreted as the average deflection 
between the source and the observer, i.e., IGMF + GMF, the Auger result suggests that 
the IGMF has low intensity and, consequently, large filling factor -- at least inside the nearby 
universe (see Figure \ref{IGMFUpperLimit}). This is indeed the most favorable scenario for individual source identification by JEM-EUSO.

\section{Individual source spectra} \label{sec:IndivSrcSpectra}

\begin{figure}[!h]
\centering
t\includegraphics [width=0.48\textwidth]{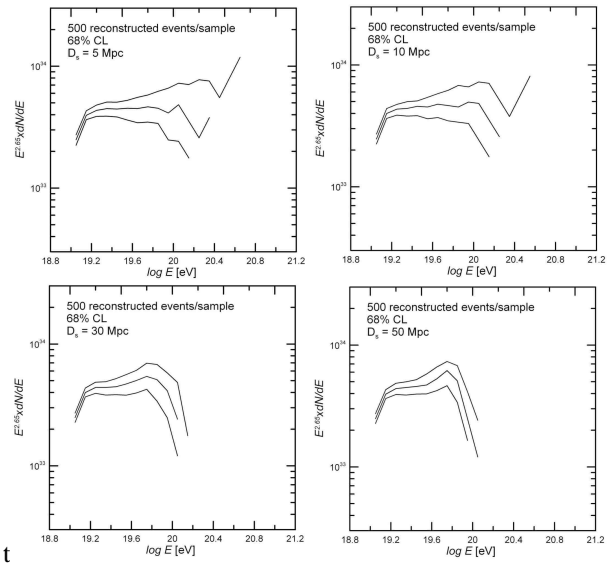}
\caption{Spectrum of a single source at distances $d=5, 10, 30$ and $50$ Mpc. The flux is kept 
constant, i.e., the intrinsic luminosity increases with $d^{2}$, and 500 events at $E>10^{19}$ eV are used in each figure. The injected spectrum is $\propto E^{-2.65}$ and particles loss energy by photo-pair and photo-meson production while traveling through a random IGMF characterized by $B = 2$ nG and $L_{c}= 300$ kpc.}
\label{SingleSrcSpectra}
\end{figure}

Given the high statistics available, the observation of large multiplicity clusters of events
will allow the evaluation of single source spectra. This is shown in Figure \ref{SingleSrcSpectra},
where the spectrum of a point source located at distances of $d=5, 10, 30$ and $50$ Mpc and 
injection spectrum $\propto E^{-2.65}$ is evaluated at the detector after propagation through IGMF
characterized by a correlation length $L_{c} = 300$ kpc and $B\sim 2 \times 10^{-9}$ G and a negligible GMF \cite{GMTancoPeVToZev}. 
Statistics are kept constant at 500 detected events per cluster at $E > 10^{19}$ eV 
(i.e. the luminosity varies with $d^{2}$ while shifting the source) and the median and the $68$\% 
confidence levels for the energy spectrum are given.

It can be seen from Figure \ref{SingleSrcSpectra} that the shape of the GZK feature can be
clearly determined in each case. This measurement gives at least two invaluable pieces of information: 
an indication of the true distance scale to the sources and a unique insight into the nature
of the high energy flux suppression, i.e., whether it is the GZK spectral complex or an acceleration
cut-off at the sources.

It is important to note that the GMF has been neglected in the calculation of the spectra in figure \ref{SingleSrcSpectra}. This is acceptable in the sense that energy losses are negligible inside the Galaxy regardless of the bending introduced by the GMF, but only as long as all the particles coming from the source, or an energy independent sample of them, is recovered at the detector. The latter is, nevertheless, a point that requires some care. 

\begin{figure}[!h]
\centering
\includegraphics [width=0.48\textwidth]{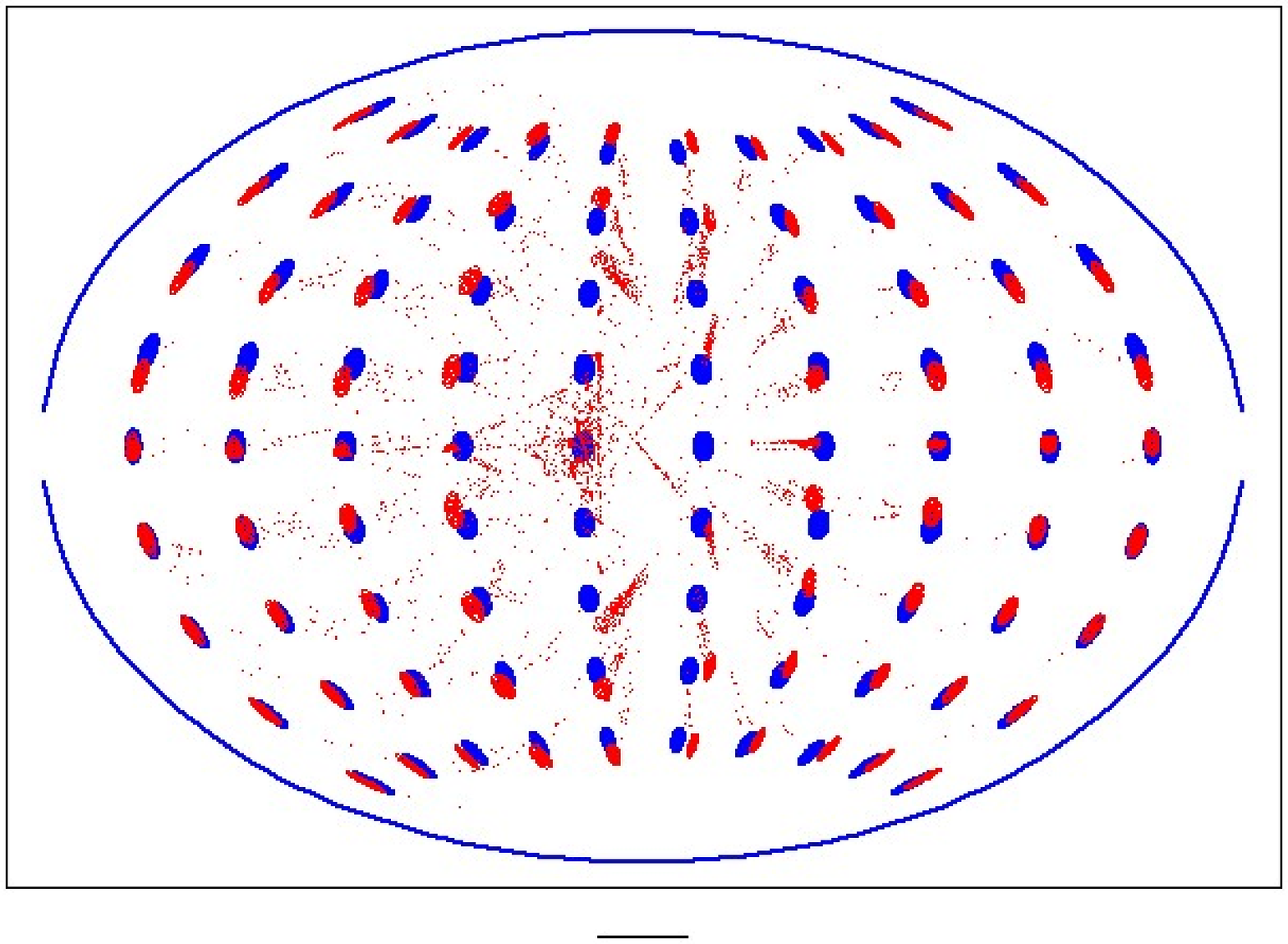}
\caption{Deflection due to the GMF. The blue circles regularly spaced in $(l,b)$ coordinates correspond to a $2.5$ deg diameter spot of $10^{20}$ eV particles impinging the external border of a spherical Halo. The red points are individual cosmic rays from those spots mapped onto the sky seen by a detector at Earth.}
\label{circles}
\end{figure}

The structure of the GMF is uncertain, especially as we move away from the Galactic plane into the Halo. However, an acceptable description can probably be made in terms of a spiral BSS thin disk embedded in an ASS Halo with a dipolar component rooted at the Galactic center and extending up to a galactocentric distance $r=20$ kpc (\cite{Han2001}). Figure \ref{circles} shows two sets of spots. The blue set is regularly spaced over a grid momentum space in galactic coordinates (pointing directions) and each one of them corresponds concentric circles of radius $0.5, 1.0, 1.5, 2.0$ and $2.5$ deg of $10^{20}$ eV protons impinging the external border of the Halo. The red dots correspond to the mapping of these UHECR onto the sky seen by the detector. It can be seen that, even at energies as high as $10^{20}$ eV deflections can be important depending on the region under consideration on the celestial sphere. That is specially true as one considers patches of sky nearer the central regions of the Galaxy. 
It must also be noted that the spots are both angularly shifted and distorted. These effects of the GMF are of course more intense as particles of lower energy are considered complicating the picture for UHECR sources that are certainly not monoenergetic.

\begin{figure}[!h]
\centering
\includegraphics [width=0.48\textwidth]{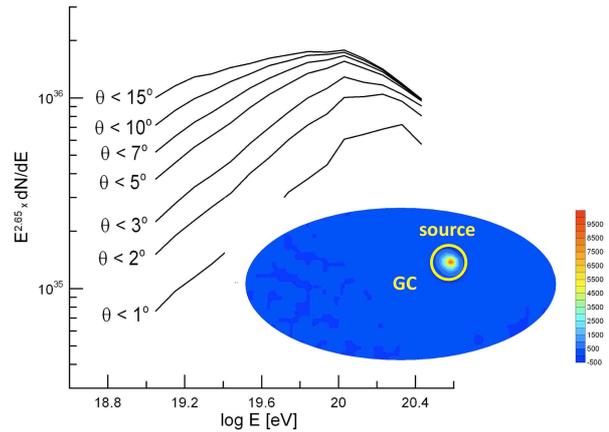}
\caption{Inset: Aitoff projection of the PSF in Galactic coordinates $(l,b)$ of a hypothetical source located at $d=5 Mpc$ for the same IGMF and injection spectrum as in figure \ref{circles}. Left: extracted spectrum as a function of the radious of a circular mask centered on the PSF used to extract the data.}
\label{PSFSpectrumHalo}
\end{figure}

Even if a conservative scenario for the IGMF is assumed in which the field has a large filling factor and $\sim $ nG average intensities, the resultant dispersion of the point spread function (PSF) of a point source may complicate the determination of the source spectrum and must be properly taken into account. Figure \ref{PSFSpectrumHalo} exemplifies this point.  The Aitoff projection of the PSF in Galactic coordinates $(l,b)$ is shown in the inset for a hypothetical source located at $d=5 Mpc$. The same IGMF and injection spectrum as in figure \ref{circles} is used. In order to isolate the effect of the IGMF, the GMF has been neglected in the calculation of this figure. The spectra obtained by selecting events inside circles centered at the PSF (which in this particular realization of the IGMF is coincident with the true position of the source, something that, in general, is not necessarily be true) are shown at the right in the same figure. It can be seen that the slope of the spectrum strongly depends on the search radius and that relevant information is scattered at distances of several degrees from the center of the PSF. This implies that a reliable estimation of the background must be implemented at the vicinity of the sources which is not a trivial requirement.

\begin{figure}[!h]
\centering
\includegraphics [width=0.48\textwidth]{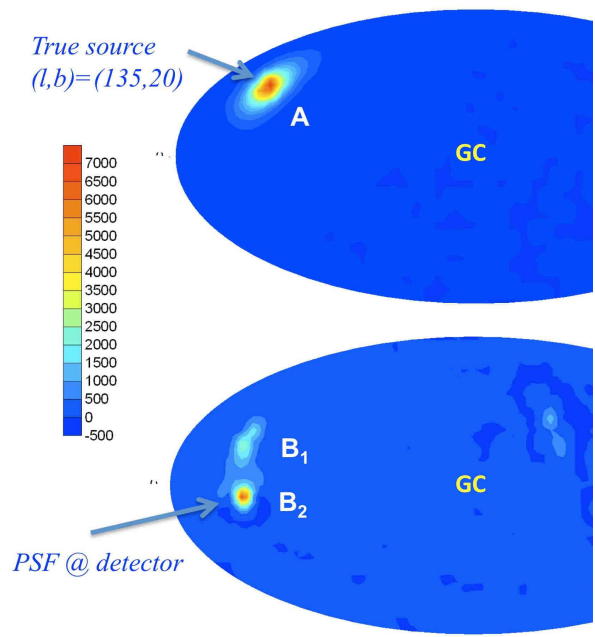}
\caption{Aitoff projection of the PSF in Galactic coordinates of a hypothetical source located at $(l,b)=(135,20)$ and $d=5 Mpc$ for the same IGMF and injection spectrum as in figure \ref{circles}. Top: at the external border of the Halo. Bottom: at the detector, after traversing the GMF. }
\label{PSFSrcHaloDetector}
\end{figure}

The problem gets more complicated when the non-negligible GMF is taken into account too in a consistent way. Figure  \ref{PSFSrcHaloDetector} shows how the PSF at the halo, A, of a given source is shifted and distorted in traversing the GMF to produce the irregular PSF B at an Earthly detector. 
Actually, the PSF B can be separated, specially if not high enough statistics are available, in two main components, $B_{1}$ and $B_{2}$. $B_{2}$ is dominated by lower energy particles and is correspondingly shifted the most up to a point that it would be difficult a priori to correlate it with the original source. $B_{1}$, on the other hand comprises more energetic particles and is only slightly shifted from the original position of the PSF prior to traversing the GMF. 

\begin{figure}[!h]
\centering
\includegraphics [width=0.48\textwidth]{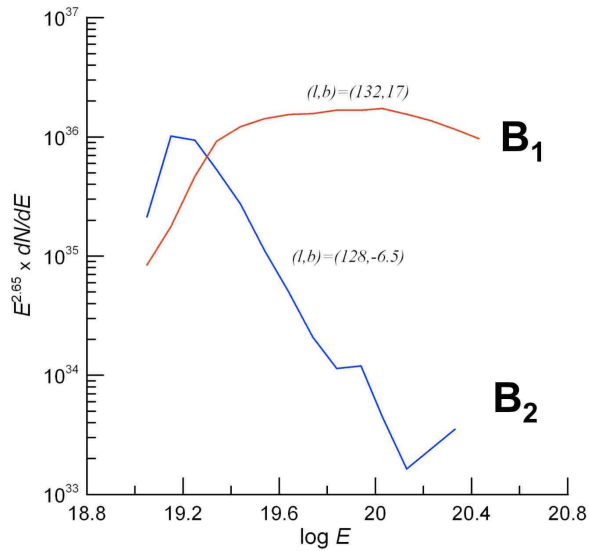}
\caption{Energy spectra extracted from the spots $B_{1}$ and $B_{2}$ of the PSF at the detector of source $A$ (see figure \ref{PSFSrcHaloDetector}).}
\label{PSFSpetrumDetector}
\end{figure}

Figure \ref{PSFSpetrumDetector} shows the spectra reconstructed from the components $B_{1}$ and $B_{2}$ respectively. It can be clearly seen that the correct spectrum may be extracted from component $B_{1}$, the nearest to the true position of the source in the sky, while the component $B_{2}$ has a very soft spectrum dominated by just the lowest energy particles from the source. Therefore, even if multiple images can be produced by the GMF for a point source, spectral measurements should help to discriminate the real image from the fake ones. Additionally, since luckily the image with the correct spectrum is angularly near to the source, high statistics and complementary information from observations at other energy bands could help pinpoint the actual astrophysical object
responsible for the acceleration of the particles.

\section{Conclusions}
 
With the advent of JEM-EUSO, the UHECR area is entering into its {\it astronomical} phase, in the sense  that individual sources will be most likely observed and analyzed for the first time. Together with that step, challenges that are common place in astronomical problems, but were foreign to high energy particle detection, like PSF determination and background treatment for image processing, will become increasingly important. This will bring an entanglement of new uncertainties but also a wealth of astrophysical information never seen before in the field. The required new set of tools is being developed at the JEM-EUSO collaboration.

\section*{Acknowledgements}
This work is partially supported by the Mexican agencies CONACyT and UNAM's CIC and PAPIIT.

\end{document}